\documentstyle[12pt]{article}

\newcommand{\vek}[1]{\mbox{{\boldmath $#1$}}}

\begin{document}

\begin{titlepage}

\title{Convolution model for the structure functions of the nucleon}

\author{J. Keppler $^{a}$, H.M. Hofmann $^{a,b}$ \\
\small \em $^{a}$Institut f\"ur theoretische Physik III, Universit\"at
Erlangen--N\"urnberg, \\
\small \em Staudtstra\ss e 7, 91058 Erlangen, Germany,  \\
\small \em email: keppler@faupt31.physik.uni--erlangen.de \\[0.3cm]
\small \em $^{b}$European Center for Theoretical Studies
in Nuclear Physics \\
\small \em and Related Areas $(ECT^{*})$, \\
\small \em Villa Tambosi, Via delle Tabarelle 286,
I--38050 Villazzano (Trento), Italy}

\maketitle

\thispagestyle{empty}

\begin{abstract}

We start from an MIT-bag model calculation
which provides information about the constituent quark distributions
in the nucleon. The constituent quarks,
however, are themselves considered as complex objects whose partonic
substructure is resolved in deep inelastic scattering.
This gives rise to structure functions of the constituent
quarks which, in the unpolarized case,
are fitted to data at a fixed scale employing three model parameters.
Using $Q^{2}$--evolution equations the data
are also well described at other scales. For the spin\-dependent
struc\-ture func\-tions $g_{1}^{p,n}$ we
additionally have to introduce polar\-izat\-ion functions for valence and sea
quarks which are determined by exploiting
the $x$--dependence of the available proton data only. A negatively
polarized sea in the range $x\geq 0.01$ is
suggested. We are then capable of predicting the shape of the neutron
structure function $g_{1}^{n}$ which turns out to be in good agreement
with experiment. Finally we present an estimate for the
trans\-versely polarized
structure function $g_{2}$, offering the possibility of extrac\-ting the
twist--3 con\-tri\-bution and rating its importance.

\end{abstract}

\end{titlepage}

\section{Introduction}

During the last two decades the extensive performance of high energy
experiments has made an important
contribution to our understanding of the
nucleon sub\-struc\-ture.
In particular deep inelastic scattering (DIS) of
leptons from nucleon targets enables the determination of the nucleon
structure functions which contain basic
information about the quark and gluon
dynamics in hadronic matter. In order to study this dynamics on the
theoretical side, one should actually
exploit QCD as the underlying theory
of strong interaction physics
which, however, is hardly understood in the
nonperturbative domain, thus offering no
possibility of calculating structure
functions from first principles. For this
reason one is forced to utilize
phenomenological approaches. But even
then we face the problem that all
available models are formulated in terms of
few quark degrees of freedom
and therefore tailored to reproduce the
low energy properties of the nucleon,
whereas for the evaluation of realistic
structure functions the nucleon
wave function should reflect the dynamics
of a large number of pointlike
partons resolved in DIS. Accordingly the
structure functions obtained in such
models have to be associated with a low
momentum scale and one has to think
about procedures to bring them in relation
to reality, i.e. the experimental
situation at $Q_{\scriptstyle\rm exp}^{2}\gg 1\,\mbox{GeV}^{2}$.

In this work we employ the traditional MIT-bag model \cite{Cho74/1} to
explicitly calculate structure functions in the low--$Q^{2}$ domain
\cite{Jaf75,Hug77,Bar79}. The picture we then
have in mind is to regard the
bag quarks as effective objects, in the subsequent sections called
{\em constituent quarks\/}, which reveal their
complicated substructure in DIS. This gives
rise to distribution functions
characterizing the dynamics of the quark
partons and gluons building up these effective quarks \cite{Alt74}.
Later in the discussion these distribution functions are denoted by
$\phi_{v},\phi_{s}$ and $\phi_{g}$. While the
constituent quark distributions
obtained in the bag model provide the
nonperturbative input of the structure
functions, the parton distributions inside the constituent quarks are
connected with large $Q^{2}$ as involved in DIS.
In this sense the constituent
quarks can be viewed as a bridge between the nonperturbative and the
perturbative regime, leading to the
convolution model for the substructure
of the nucleon \cite{Hwa80}.

The first goal of our investigations is to
demonstrate that a satisfactory and
consistent description of the unpolarized
structure functions can be achieved
in such a convolution model approach.
In this framework the $Q^{2}$--evolution
equations can be completely expressed
in terms of the parton distribution
functions $\phi$, exposing the change of
the substructure of the constituent
quarks with increasing resolution. We are
then going to apply the formalism to
the longitudinally polarized structure functions of the nucleon, i.e.
$g_{1}^{p,n}$, where a careful analysis of the latest data
\cite{EMC89,SMC94} is presented. In doing this, we solely exploit the
$x$--dependence of $g_{1}^{p}$ and use
neither the extrapolated integrals
over the data nor information from hyperon $\beta$--decays. Our special
attention in this analysis is given to the
polarization of the sea quarks
which turns out to be essential for the
understanding of $g_{1}$. With the
sea polarization in hands a prediction
for the neutron structure function
$g_{1}^{n}$ can be made and compared with the available data
\cite{SLAC93,SMC93}. Finally we will also present an estimate for the
structure function $g_{2}$ which will be experimentally accessible via
transversely polarized scattering
events \cite{Prop1,Prop2} and for the first
time offers the possibility of extracting
higher--twist contributions in DIS
\cite{Jaf90}.

The paper is organized as follows.
In Sect.\ 2 we will start with a brief
review of the DIS formalism and the definitions of structure functions.
Sect.\ 3  deals with the interpretation of
structure functions resulting from
quark model calculations. This leads to the concept of the convolution
model, as explained in Sect.\ 4. Thereupon
we will successively devote us to
the discussion of $F_{2}$, $g_{1}$, and $g_{2}$, corresponding to
Sects.\ 5, 6, and 7, respectively. In Sect.\ 8
we give a conclusion of our
results as well as an outlook on problems which remain to be solved.

\section{Deep inelastic scattering and structure functions}

The Feynman diagram which describes the deep inelastic lepton--nucleon
scattering process $\ell + N \rightarrow \ell' + X$ in
leading order of the electromagnetic interaction is depicted in
Fig.\ 1.
For further considerations we go into the target rest frame in which
\begin{equation}\label{Eq1}
p=(M,0)\;,\hspace{0.7cm} k=(E,\vek{k})\;,\hspace{0.7cm} k'=(E',\vek{k}')
\end{equation}
holds and write the frequently appearing quantities as
\begin{eqnarray}
\nu & = & \frac{p\cdot q}{M} = E-E'\;, \label{Eq2}   \\
Q^{2} & = & -q^{2}\;, \label{Eq3}     \\
x & = & \frac{Q^{2}}{2\,p\cdot q} = \frac{Q^{2}}{2M\nu}\;.  \label{Eq4}
\end{eqnarray}
Employing the usual Feynman rules to evaluate the inclusive differential
cross section, one finds \cite{Clo79}
\begin{equation}\label{Eq5}
\frac{d^{2}\sigma}{dE'\,d\Omega_{\ell}} = \frac{\alpha^{2}}{Q^{4}}
\frac{E'}{E}L^{\mu\nu}W_{\mu\nu}\;.
\end{equation}
While the leptonic tensor $L^{\mu\nu}$ can
be calculated explicitly due to
the pointlike nature of the
lepton, the hadronic tensor $W_{\mu\nu}$
contains all the relevant information
about the substructure of the nucleon
and therefore represents a much more
complicated object. For the spin
averaged case we get two completely symmetric tensors \cite{CL88}
\begin{eqnarray}
L_{S}^{\mu\nu} & = & \frac{1}{2} \sum_{s,s'}
\left[\bar{u}(k',s')\gamma^{\mu} u(k,s)\right]
\left[\bar{u}(k',s')\gamma^{\nu} u(k,s)\right] \nonumber \\
& = & \frac{1}{2}\mbox{Tr}\left[\gamma^{\mu}(\not k+m)
\gamma^{\nu}(\not k'+m)\right]\;,  \label{Eq6} \\
W_{\mu\nu}^{S} & = & \frac{1}{4\pi M}\,\frac{1}{2} \sum_{\sigma} \int
d^{4}\xi \,e^{iq\cdot \xi}
\langle p,\sigma|[J_{\mu}(\xi),J_{\nu}(0)]^{S}|p,\sigma\rangle \;,
\label{Eq7}
\end{eqnarray}
$J_{\mu}$ being the operator of the
hadronic electromagnetic current. The
unpolarized structure functions $W_{1,2}$ are
defined by means of the most
general symmetric tensor respecting current conservation and Lorentz
invariance \cite{Clo79}
\begin{equation}\label{Eq8}
W_{\mu\nu}^{S} = W_{1}(\nu,q^{2})\left(-g_{\mu\nu}+
\frac{q_{\mu}q_{\nu}}{q^{2}}\right)+ \frac{W_{2}(\nu,q^{2})}{M^{2}}
\left(p_{\mu}-\frac{p\cdot q}{q^{2}}q_{\mu}\right)
\left(p_{\nu}-\frac{p\cdot q}{q^{2}}q_{\nu}\right)\;.
\end{equation}
At that point one usually switches over
to dimensionless structure functions
$F_{1,2}$ which are related to $W_{1,2}$ in the following way:
\begin{eqnarray}
F_{1}(x,Q^{2}) & = & M\, W_{1}(\nu,q^{2}) \;, \label{Eq9} \\
F_{2}(x,Q^{2}) & = & \nu\, W_{2}(\nu,q^{2}) \;. \label{Eq10}
\end{eqnarray}
In the naive quark parton model \cite{Fey72} these functions can be
expressed in terms of quark distribution functions $q_{i}^{N}$ of the
various flavours $i$ in the nucleon $N$,
\begin{equation}\label{Eq11}
F_{2}^{N}(x)=2xF_{1}^{N}(x) =
\sum_{i=1}^{2n_{f}} e_{i}^{2}\, x \, q_{i}^{N}(x)\;,
\end{equation}
where the independence on $Q^{2}$ is a consequence of the neglect of
QCD--corrections to the dominant photon--parton scattering process.

In the case of spindependent lepton--nucleon scattering the information
which is specific for the polarization effects enters the completely
antisymmetric parts of the tensors \cite{Clo79,CL88}
\begin{eqnarray}
L_{A}^{\mu\nu} & = & \frac{1}{2}\mbox{Tr}
\left[\gamma^{\mu}\gamma^{5}(\not k+m)
\gamma^{\nu}(\not k'+m)\right]\;,  \label{Eq12} \\
W_{\mu\nu}^{A} & = & \frac{1}{4\pi M} \int
d^{4}\xi \,e^{iq\cdot \xi}
\langle p,\sigma|[J_{\mu}(\xi),J_{\nu}(0)]^{A}|p,\sigma\rangle
\label{Eq13}  \\
& = &  i\,\varepsilon_{\mu\nu\alpha\rho}\, q^{\alpha}
\left[s^{\rho}M\,G_{1}+\left(s^{\rho}\frac{p\cdot q}{M}-
p^{\rho}\frac{s\cdot q}{M}\right)G_{2}\right]\;,
\label{Eq14}
\end{eqnarray}
with $s^{\rho}$ denoting the spin vector
of the nucleon. The last expression
again constitutes the most general
ansatz and thus serves as a definition for
the functions $G_{1,2}$ which are related
to the commonly used structure
functions $g_{1,2}$ by
\begin{eqnarray}
g_{1}(x,Q^{2}) & = & M^{2}\nu\, G_{1}(\nu,q^{2}) \;, \label{Eq15} \\
g_{2}(x,Q^{2}) & = & M\nu^{2}\, G_{2}(\nu,q^{2}) \;. \label{Eq16}
\end{eqnarray}
In the naive parton model one finds \cite{Fey72}
\begin{eqnarray}
g_{1}^{N}(x) & = &
\frac{1}{2}\sum_{i=1}^{2n_{f}} e_{i}^{2}
\left(q_{i}^{\uparrow \,N}(x)-q_{i}^{\downarrow \,N}(x)\right)
\equiv \frac{1}{2}\sum_{i=1}^{2n_{f}} e_{i}^{2}\,\Delta q_{i}^{N}(x),
\label{Eq17} \\
g_{T}^{N}(x) \equiv g_{1}^{N}(x)+g_{2}^{N}(x) & = &
\frac{1}{2}\sum_{i=1}^{2n_{f}} e_{i}^{2}
\left(\tilde{q}_{i}^{\uparrow \,N}(x)-
\tilde{q}_{i}^{\downarrow \,N}(x)\right)
\equiv \frac{1}{2}\sum_{i=1}^{2n_{f}} e_{i}^{2}\,
\Delta \tilde{q}_{i}^{N}(x),
\label{Eq18}
\end{eqnarray}
where $q_{i}^{\uparrow (\downarrow) \,N}$ characterize the probabilities
to find in a longitudinally polarized
nucleon a quark with flavour $i$ and
spin alignment parallel (antiparallel) to the nucleon spin, while
$\tilde{q}_{i}^{\uparrow (\downarrow) \,N}$ are the correponding
probabilities for transversely polarized nucleons.

In addition to the fundamental process $\gamma^{\ast}q\rightarrow q$ the
framework of QCD supplies further processes which lead to corrections of
the naive parton model. Taking such
modifications into account, the above
introduced quark distributions acquire
a $Q^{2}$--dependence. Furthermore
the QCD--improved formalism also involves the gluon distribution
$g(x,Q^{2})$. Adopting the variable
\begin{equation}\label{Eq19}
\kappa = \frac{2}{\beta_{0}}\ln\left(
\frac{\ln\left(\frac{Q_{0}^{2}}{\Lambda^{2}}\right)}{
\ln\left(\frac{Q^{2}}{\Lambda^{2}}\right)}\right)\equiv
-\frac{2}{\beta_{0}}\ln\zeta\;,
\end{equation}
with $\beta_{0}=11-\frac{2}{3}n_{f}$, and
using the non--singlet (NS) and
singlet (S) quark distributions
\begin{eqnarray}
q^{\scriptstyle\rm NS}(x,Q^{2}) & = & \sum_{i=1}^{n_{f}}
\biggl(q_{i}(x,Q^{2}) - \bar{q}_{i}(x,Q^{2})\biggr)\;, \label{Eq20} \\
q^{\scriptstyle\rm S}(x,Q^{2}) & = & \sum_{i=1}^{n_{f}}
\biggl(q_{i}(x,Q^{2}) + \bar{q}_{i}(x,Q^{2})\biggr)\;, \label{Eq21}
\end{eqnarray}
the leading order QCD--evolution equations finally can be written as
\cite{AP77}
\begin{eqnarray}
\frac{d}{d\kappa}\,q^{\scriptstyle\rm NS}(x,Q^{2})
& = & \int\limits_{x}^{1}
\frac{dy}{y}\,P_{q\to qg}\!\left(\frac{x}{y}\right)
q^{\scriptstyle\rm NS}(y,Q^{2})  \;, \label{Eq22} \\
\frac{d}{d\kappa}
\left(\begin{array}{c} q^{\scriptstyle\rm S}(x,Q^{2}) \\
g(x,Q^{2}) \end{array} \right)
& = & \int\limits_{x}^{1}
\frac{dy}{y}
\left(\begin{array}{cc}
P_{q\to qg}\!\left(\frac{x}{y}\right) &
2n_{f}\,P_{g\to q\bar{q}}\!\left(\frac{x}{y}\right) \\
P_{q\to gq}\!\left(\frac{x}{y}\right) &
P_{g\to gg}\!\left(\frac{x}{y}\right)
\end{array} \right)
\left(\begin{array}{c} q^{\scriptstyle\rm S}(y,Q^{2}) \\
g(y,Q^{2}) \end{array}
\right) \;, \label{Eq23}
\end{eqnarray}
$P$ being the splitting functions of the various splitting processes.
These evolution equations enable the determination of $F_{2}$ at any
perturbatively accessible scale $Q^{2}$, provided one knows the function
at a reference scale $Q_{0}^{2}$. Whereas on the one hand for the
longitudinally polarized distribution functions analogous equations hold
with $q$ and $g$ replaced by $\Delta q$ and $\Delta g$, respectively, a
careful analysis of the operator structure of $g_{2}$ \cite{JX91} on the
other hand reveals that $g_{2}$ also receives twist--3 contributions
so that its $Q^{2}$--evolution remains a yet unsolved problem.

\section{Interpretation of quark model calculations}

So far the employment of the parton model merely resulted in a
parametrization of the structure
functions in terms of quark distributions
which contain nonperturbative information
about the hadronic bound state. From the
theoretical point of view it must be the goal to calculate the
structure functions, and with those also the quark distributions,
using appropriate models of the nucleon.
Basically this is possible with the help of Eqs.\
(\ref{Eq7}),(\ref{Eq8}) and (\ref{Eq13}),(\ref{Eq14})
after having performed
an operator product expansion of the
current commutator. Equipped with this
formalism the final step towards the explicit calculation of structure
functions consists in inserting a nucleon wave function into the matrix
elements of (\ref{Eq7}) and (\ref{Eq13}).
In order to get realistic
results, this wave function should be formulated in terms of pointlike
partons which are resolved in DIS.
In practice, however, there is so far no
knowledge whatsoever how to transform the dynamics of the
elementary constituents of the nucleon
into a wave function. This is why
one is for the presence forced to
utilize well established, phenomenological
low energy models of the nucleon
which are based on few quark degrees of
freedom. One of these models is the
MIT-bag model \cite{Cho74/1} which is
formulated relativistically and
incorporates confinement in a simple and
transparent manner, thus being very well suited for the calculation of
structure functions. Using the
well known $SU(6)$ nucleon wave functions,
the results for the physical structure functions in the bag model read
as follows \cite{Jaf75,Hug77,Bar79}
\begin{eqnarray}
\biggl( F_{1}^{p}(x)\biggr)_{\scriptstyle\rm bag}
& = & \frac{MR}{2\pi}\frac{\varepsilon^{4}}
                          {\varepsilon^{2}-\sin^{2}\varepsilon}
\Biggl\{ \int\limits_{|MR\,x-\varepsilon|}^{\infty}d\beta\,\beta
\Biggl[ T_{0}^{2}(\varepsilon,\beta) + T_{1}^{2}(\varepsilon,\beta)
\nonumber \\
& & \qquad -\frac{2}{\beta}\,(\varepsilon-MR\,x)\,
T_{0}(\varepsilon,\beta)\,T_{1}(\varepsilon,\beta) \Biggr]
\; + \; (x\rightarrow -x) \Biggr\}\;,  \label{Eq24}  \\
\biggl( g_{1}^{p}(x)\biggr)_{\scriptstyle\rm bag}
& = & \frac{5}{9}\frac{MR}{2\pi}\frac{\varepsilon^{4}}
                                {\varepsilon^{2}-\sin^{2}\varepsilon}
\Biggl\{ \int\limits_{|MR\,x-\varepsilon|}^{\infty}d\beta\,\beta
\Biggl[ T_{0}^{2}(\varepsilon,\beta) \nonumber \\
& & \qquad + \left( 2\left( \frac{\varepsilon-MR\,x}{\beta}
\right)^{\!\! 2} -1 \right)
T_{1}^{2}(\varepsilon,\beta) \nonumber \\
& & \qquad -\frac{2}{\beta}\,(\varepsilon-MR\,x)\,
T_{0}(\varepsilon,\beta)\,T_{1}(\varepsilon,\beta) \Biggr]
\; + \; (x\rightarrow -x) \Biggr\}\;,  \label{Eq25}  \\
\biggl( g_{T}^{p}(x)\biggr)_{\scriptstyle\rm bag}
& = & \frac{5}{9}\frac{MR}{2\pi}\frac{\varepsilon^{4}}
                                {\varepsilon^{2}-\sin^{2}\varepsilon}
\Biggl\{ \int\limits_{|MR\,x-\varepsilon|}^{\infty}d\beta\,\beta
\Biggl[ T_{0}^{2}(\varepsilon,\beta)  \nonumber \\
& & \qquad -\left( \frac{\varepsilon-MR\,x}{\beta} \right)^{\!\! 2}
T_{1}^{2}(\varepsilon,\beta) \Biggr]
\; + \; (x\rightarrow -x) \Biggr\}\;,  \label{Eq26}
\end{eqnarray}
with
\begin{equation}\label{Eq27}
T_{n}(\varepsilon,\beta) = \int\limits_{0}^{1}dz\,z^{2}\,
j_{n}(\varepsilon z)\, j_{n}(\beta z)\;.
\end{equation}
In the case of the neutron we have
$\left( F_{1}^{n}(x)\right)_{\scriptstyle\rm bag} =
\frac{2}{3}\left( F_{1}^{p}(x)\right)_{\scriptstyle\rm bag}$,
wheras both
$\left( g_{1}^{n}(x)\right)_{\scriptstyle\rm bag}$ and
$\left( g_{T}^{n}(x)\right)_{\scriptstyle\rm bag}$ vanish identically.
The quantities $\varepsilon$ and $MR$ entering the expressions for the
structure functions denote the lowest bag frequency and the product
mass times bag radius, respectively, and are
completely fixed by the boundary
and stability condition of the bag, giving $\varepsilon=2.04$ and
$MR=4\varepsilon$ \cite{Cho74/2}, so
that Eqs.\ (\ref{Eq24})--(\ref{Eq26})
do not contain any free parameter.
As a consequence of broken translational
invariance in the static approximation of the bag model the
structure functions receive
nonvanishing support from the unphysical region
$x>1$. For this reason modified versions
of the bag model \cite{BM87,SST91} are equipped with
a Peierls--Yoccoz projection \cite{PY57} onto
momentum eigenstates
which weakens but not completely cures
the support problem. In this work
we will disregard such a projection since we consider the latter as a
minor deficiency. In fact, the numerical contribution from the region
$x>1$ is very small and the normalizations
\begin{eqnarray}
\int\limits_{0}^{1} dx
\biggl( F_{1}^{p}(x)\biggr)_{\scriptstyle\rm bag}
& = & \frac{1}{2}\;,  \label{Norma}   \\
\int\limits_{0}^{1} dx \biggl( g_{1}^{p}(x)\biggr)_{\scriptstyle\rm bag}
& = & \frac{1}{6}\left(\frac{g_{A}}{g_{V}}\right)
_{\scriptstyle\rm bag}\;,  \label{Normb}  \\
\int\limits_{0}^{1} dx \biggl( g_{2}^{p}(x)\biggr)_{\scriptstyle\rm bag}
& = & 0  \label{Normc}
\end{eqnarray}
are fulfilled in good approximation. Here
$(g_{A}/g_{V})_{\scriptstyle\rm bag} = 1.09$ denotes
the ratio of the weak
coupling constants in the bag model
which underestimates the experimental
one by roughly $12\%$ \cite{Cho74/2}.

The results for the structure functions summarized in the previous
paragraph were obtained with the help of a wave function which is
formulated in terms of three
bag quarks. The dynamics of these degrees of
freedom yields a fair description
of the low energy properties of the nucleon
which are characterized by observables
such as mass, magnetic moment, weak
coupling constants, charge radius, and
electromagnetic form factors. All
these quantities are associated with a very low momentum scale
$Q^{2}\ll 1\,\mbox{GeV}^{2}$.
Comparing, however, the structure functions
derived from the nucleon wave function
of the bag model with experiment,
{\em no\/} agreement can be seen.
As already mentioned before, this of course
is due to the fact that the structure functions measured in DIS are
associated with very large momentum scales
$Q^{2}\gg 1\,\mbox{GeV}^{2}$ implying a high resolution of the nucleon
substructure. In order to bring then the structure functions of the bag
model in relation to reality, we regard the quarks of the bag model as
effective degrees of freedom. To be precise,
we understand these effective
quarks as clusters which in addition
to the valence quarks also consist of
sea quarks and gluons \cite{Hwa80}.
The effective quarks defined in this
sense are designated as {\em constituent quarks\/} in the following.
Accordingly the structure functions calculated in the framework of the
bag model, Eqs.\ (\ref{Eq24})--(\ref{Eq26}),
can be interpreted as linear
combinations of constituent
quark distributions, labeled by the index $C$.
In analogy to Eqs.\ (\ref{Eq11}), (\ref{Eq17})
and (\ref{Eq18}) we then
get
\begin{eqnarray}
\frac{1}{2x} \biggl( F_{2}^{N}(x)\biggr)_{\scriptstyle\rm bag} =
\biggl( F_{1}^{N}(x)\biggr)_{\scriptstyle\rm bag}
& = & \frac{1}{2}\sum_{C} e_{C}^{2}\,q_{C}^{N}(x)\;,
\label{Eq28} \\
\biggl( g_{1}^{N}(x)\biggr)_{\scriptstyle\rm bag}
& = & \frac{1}{2} \sum_{C} e_{C}^{2}\,\Delta q_{C}^{N}(x)\;,
\label{Eq29} \\
\biggl( g_{T}^{N}(x)\biggr)_{\scriptstyle\rm bag}
& = & \frac{1}{2}\sum_{C} e_{C}^{2}\,
\Delta \tilde{q}_{C}^{N}(x)\;.
\label{Eq30}
\end{eqnarray}

\section{The convolution model}

The concept of the constituent quarks serves as a bridge between the
nonperturbative and perturbative regime.
At low $Q^{2}$--scales the resolution
is very poor and the hadron appears
to be a bound state of three constituent
quarks which seem to be structureless. In reality, however, they have a
complicated substructure which is
resolved in DIS. On the formal level this
picture can be translated into a corresponding diagram \cite{Jaf85}
depicted in Fig.\ 2. It describes a nucleon $N$ with momentum
$p$ containing a constituent quark $C$ with momentum $k$ which in turn
consists of a quark parton of flavour $i$ carrying momentum $\tilde{k}$.
This parton absorbs the virtual photon. Fig.\ 2
is a reflection of an impulse approximation which implies the
assumptions that final state interactions on the one hand between the
fragments of the nucleon and the constituent quark and on the other hand
between the
fragments of the constituent quark and
the parton are neglected. While the
latter are suppresed by $1/Q^{2}$,
as in the parton model, there is, however,
no argument for the suppression of the former
because of the absence of a
characteristic mass scale. Nevertheless,
in the weak coupling limit of the
constituent quarks, as realized in the bag model,
the impulse approximation
should be justified very well \cite{Jaf85}.

Taking the momentum dependences of the
quark distributions, as given in
Fig.\ 2, into consideration, the probability of finding a quark $i$
in the nucleon $N$ is determined by
\begin{equation} \label{Eq31}
q_{i}^{N}(p,q) = \sum_{C}
\int d^{4}k\, q_{C}^{N}(p,k) \, q_{i}^{C}(k,q)\;.
\end{equation}
After transition to the usual invariants one finally arrives at
\cite{Jaf85}
\begin{equation} \label{Eq32}
q_{i}^{N}(x,Q^{2}) = \sum_{C} \int \limits_{x}^{1}\frac{dy}{y}
\, q_{C}^{N}(y) \: q_{i}^{C}\!\!\left( \frac{x}{y},Q^{2} \right) \;.
\end{equation}
This is the fundamental relation of the
convolution model which describes
the parton distributions in the nucleon
as a convolution of the constituent
quark distributions in the nucleon and the parton distributions in the
constituent quarks.

Let us now have a closer look at the parton distributions in the
constituent quarks. Since the formalism has to include valence as
well as sea quarks, it is advantageous to
decompose the functions $q_{i}^{C}$
into two corresponding parts \cite{Dzi89},
\begin{equation} \label{Eq33}
q_{i}^{C}(z,Q^{2}) = \phi_{v}(z,Q^{2})\,\delta_{iC} +
\phi_{s}^{i/C}(z,Q^{2})\;.
\end{equation}
While the sea quark distributions have
to be characterized by an isospin label
in general, the valence quark distribution
should be isospin independent
and has to fulfill the normalization constraint
\begin{equation} \label{Eq34}
\int \limits_{0}^{1} dz\, \phi_{v}(z,Q^{2}) = 1 \;.
\end{equation}
This together with
the $\delta_{iC}$ of Eq.\ (\ref{Eq33}) is a reflection
of associating with each constituent quark just one valence quark of
the same flavour. For the sea quark distributions it is reasonable to
impose isospin symmetry with
respect to the constituent quarks as well as
full symmetry of the strange quarks, yielding finally
the most general relations
\begin{eqnarray}
\phi_{s}^{u/U}(z,Q^{2}) & = & \phi_{s}^{\bar{u}/U}(z,Q^{2})
= \phi_{s}^{d/D}(z,Q^{2}) = \phi_{s}^{\bar{d}/D}(z,Q^{2}) \;,
\label{Eq35} \\
\phi_{s}^{u/D}(z,Q^{2}) & = & \phi_{s}^{\bar{u}/D}(z,Q^{2})
= \phi_{s}^{d/U}(z,Q^{2}) = \phi_{s}^{\bar{d}/U}(z,Q^{2}) \;,
\label{Eq36} \\
\phi_{s}^{s/U}(z,Q^{2}) & = & \phi_{s}^{\bar{s}/U}(z,Q^{2})
= \phi_{s}^{s/D}(z,Q^{2}) = \phi_{s}^{\bar{s}/D}(z,Q^{2}) \;.
\label{Eq37}
\end{eqnarray}

\section{The unpolarized structure function $F_{2}$}

Based on the relations of the previous section we are now capable of
formulating the convolution model
expressions for the unpolarized structure
function $F_{2}$. In addition to the constituent quark distributions
$q_{C}^{N}$ this expression also involves the
parton distributions $\phi$.
The former reflect the dynamics of constituent quarks inside hadrons
and contain information about the
confinement mechanism. In the following
this nonperturbative part of $F_{2}$
will be described by the bag model
calculation (\ref{Eq24}). The latter
contain information about the parton
dynamics in the constituent quarks.
Since, however, there is no way to
calculate these functions $\phi$ at a
microscopic level, one has to use
suitable parametrizations at a reference scale $Q_{0}^{2}$. In order to
minimize the number of free parameters, we start our considerations
by equating (\ref{Eq35})
and (\ref{Eq36}), i.e. $\phi_{s}^{u/U}(z,Q^{2})
= \phi_{s}^{u/D}(z,Q^{2})\equiv \phi_{s}(z,Q^{2})$ and relating
(\ref{Eq37}) to (\ref{Eq35}) by the $z$--independent suppression factor
$1/2$ \cite{Rob90},
i.e. $\phi_{s}^{s/U}(z,Q^{2}) = \frac{1}{2}\phi_{s}^{u/U}(z,Q^{2})$.
Bearing the normalization
condition (\ref{Eq34}) in mind, we then choose
\cite{Alt74}
\begin{eqnarray}
\phi_{v}(z,Q_{0}^{2}) & = & \frac{\Gamma (a+3/2)}
                                 {\Gamma (1/2)\,\Gamma (a+1)}
\frac{1}{\sqrt{z}}(1-z)^{a} \;, \label{Eq38} \\
\phi_{s}(z,Q_{0}^{2}) & = &
A\,\frac{1}{z}(1-z)^{b} \;. \label{Eq39}
\end{eqnarray}
These expressions are motivated by Regge arguments and dimensional
counting rules. Employing (\ref{Eq11}), (\ref{Eq28}),
(\ref{Eq32}), (\ref{Eq33}) and $SU(6)$--symmetry, we finally get
\begin{equation} \label{Eq40}
F_{2}^{p,n}(x,Q^{2}) = 2x \int\limits_{x}^{1} \frac{dy}{y}
\biggl( F_{1}^{p,n}(y)\biggr)_{\scriptstyle\rm bag} \phi_{v}
\!\!\left( \frac{x}{y},Q^{2} \right) + \frac{22}{3}x
\int\limits_{x}^{1} \frac{dy}{y}
\biggl( F_{1}^{p}(y)\biggr)_{\scriptstyle\rm bag} \phi_{s}
\!\!\left( \frac{x}{y},Q^{2} \right) \;.
\end{equation}
The total of three parameters
entering $\phi_{v}$ and $\phi_{s}$ are fitted
to experiment at the reference scale $Q_{0}^{2} = 10\,\mbox{GeV}^{2}$,
giving $a=-0.17$, $b=4$, and $A=0.137$. Fig.\ 3 shows the result
in comparison with the data \cite{BCDMS-F2}.

With the help of the QCD--evolution equations
(\ref{Eq22}) and (\ref{Eq23}) we can then
determine the structure function
at any other scale $Q^{2}$ which lies in the perturbatively accessible
domain. This evolution also involves the gluon distribution which in the
convolution model can be written as
\begin{equation} \label{Eq41}
g(x,Q^{2}) = 6 \int\limits_{x}^{1} \frac{dy}{y}
\biggl( F_{1}^{p}(x)\biggr)_{\scriptstyle\rm bag} \phi_{g}
\!\!\left( \frac{x}{y},Q^{2} \right)\;.
\end{equation}
Since the total $Q^{2}$--dependence of $F_{2}$ is fully contained
in the functions $\phi$, it is desirable to derive evolution equations
for them. Performing the transition to the moments
\begin{eqnarray}
\left( \begin{array}{c}
M_{n}^{\phi_{v}}(Q^{2})  \\ M_{n}^{\phi_{s}}(Q^{2}) \\
M_{n}^{\phi_{g}}(Q^{2}) \end{array} \right) & = &
\int\limits_{0}^{1}dz\,z^{n-1}
\left( \begin{array}{c}
\phi_{v}(z,Q^{2})  \\ \phi_{s}(z,Q^{2}) \\
\phi_{g}(z,Q^{2}) \end{array} \right) \;, \label{Eq42}  \\
\left( \begin{array}{c}
A_{n}^{\scriptstyle\rm NS}  \\ A_{n}^{gq} \\
A_{n}^{q\bar{q}} \\ A_{n}^{gg} \end{array} \right) & = &
\int\limits_{0}^{1}dz\,z^{n-1}
\left( \begin{array}{c}
P_{q\to qg}(z)  \\ P_{q\to gq}(z) \\
P_{g\to q\bar{q}}(z)  \\ P_{g\to gg}(z)\end{array} \right) \;,
\label{Eq43}
\end{eqnarray}
\begin{equation} \label{Eq44}
A_{n}^{\pm} = \frac{1}{2}\left( A_{n}^{\scriptstyle\rm NS} +
A_{n}^{gg} \pm \sqrt{
                \left( A_{n}^{\scriptstyle\rm NS}
                - A_{n}^{gg} \right)^{\!2}
                 +4\, A_{n}^{gq}\, 2n_{f}\, A_{n}^{q\bar{q}} }
\right) \;,
\end{equation}
and taking advantage of the properties of convolution integrals, we
finally get the solutions
\begin{eqnarray}
M_{n}^{\phi_{v}}(Q^{2}) & = &
\zeta^{-\frac{2}{\beta_{0}} A_{n}^{\scriptscriptstyle\rm NS} }
M_{n}^{\phi_{v}}(Q_{0}^{2}) \;, \label{Eq45}  \\
M_{n}^{\phi_{s}}(Q^{2}) & = &
\frac{1}{5} \Biggl\{ \frac{1}{A_{n}^{-} - A_{n}^{+}} \Biggl\{ \Bigl[
\left( A_{n}^{-} - A_{n}^{\scriptscriptstyle\rm NS} \right)
\zeta^{ -\frac{2}{\beta_{0}} A_{n}^{+} } +
\left( A_{n}^{\scriptscriptstyle\rm NS} - A_{n}^{+}\right)
\zeta^{ -\frac{2}{\beta_{0}} A_{n}^{-} } \Bigr] \times \nonumber \\
& & \hspace{5.6cm} \times \left(
M_{n}^{\phi_{v}}(Q_{0}^{2}) + 5\, M_{n}^{\phi_{s}}(Q_{0}^{2}) \right)
\nonumber \\
& & \hspace{0.8cm} +2n_{f} \,A_{n}^{q\bar{q}}
\left( \zeta^{ -\frac{2}{\beta_{0}} A_{n}^{-} } -
\zeta^{ -\frac{2}{\beta_{0}} A_{n}^{+} }\right)
                    M_{n}^{\phi_{g}}(Q_{0}^{2})
\Biggr\} - M_{n}^{\phi_{v}}(Q_{0}^{2}) \Biggr\}\;. \label{Eq46}
\end{eqnarray}
The variable $\zeta$ has been defined in Eq.\ (\ref{Eq19}), explicit
expressions for the moments $A$ can
be found in \cite{AP77}. For the gluon
distribution which also cannot be
deduced by means of a microscopic model
yet, we use the parametrization
\begin{equation} \label{Eq47}
\phi_{g}(z,Q_{0}^{2}) =
B\,\frac{1}{z}(1-z)^{c} \;,
\end{equation}
with $B=1.48$, $c=2.37$. According to (\ref{Eq41})
this parameter choice corresponds to
$g(x,Q_{0}^{2}) = 4.27\frac{1}{x}(1-x)^{8}$
which respects the momentum sum rule $\int_{0}^{1}dx\,x
(q^{\scriptstyle\rm S}(x,Q^{2}) + g(x,Q^{2})) = 1$ in conjunction with
the singlet quark distribution and turns out to be a reasonable ansatz.
Finally the QCD--parameter $\Lambda$
entering $\zeta$ is settled at $\Lambda = 200\,\mbox{MeV}$.

Employing standard techniques \cite{Bur80},
we reconstruct the distribution
functions $\phi_{v}(z,Q^{2})$
and $\phi_{s}(z,Q^{2})$ from the evolved
moments, Eqs.\ (\ref{Eq45}), (\ref{Eq46}), using twelve and three
moments, respectively.
These distributions tell us in which way the
parton substructure of the constituent
quarks changes with the resolution
$Q^{2}$. The results for typical scales
are displayed in Figs.\ 4 and 5. From
these we can easily determine the $Q^{2}$--dependence of the
structure function $F_{2}^{p}$; the results
are compared to some data in Figs.\ 6 and 7.

After having restricted ourselves to the unpolarized structure function
of the proton so far, some remarks about the neutron structure function
are in order now. As usual we investigate the difference
$F_{2}^{p}-F_{2}^{n}$ which in the convolution model is given by
\begin{eqnarray}
\left[ F_{2}^{p}-F_{2}^{n} \right](x,Q^{2}) & = &
\frac{2}{3} x \int\limits_{x}^{1} \frac{dy}{y}
\biggl( F_{1}^{p}(y)\biggr)_{\scriptstyle\rm bag} \phi_{v}
\!\!\left( \frac{x}{y},Q^{2} \right) \nonumber \\
& & - \frac{4}{3}x
\int\limits_{x}^{1} \frac{dy}{y}
\biggl( F_{1}^{p}(y)\biggr)_{\scriptstyle\rm bag}
\left[ \phi_{s}^{u/D} \!\!\left( \frac{x}{y},Q^{2} \right) -
\phi_{s}^{u/U} \!\!\left( \frac{x}{y},Q^{2} \right) \right]\! \!.
\label{Eq48}
\end{eqnarray}
Assuming $\phi_{s}^{u/U} = \phi_{s}^{u/D}$, as before, the sea quark
contribution drops and we are left solely with the valence part.
Fig.\ 8 shows $F_{2}^{p}-F_{2}^{n}$ in this case for
$Q^{2}= 40\,\mbox{GeV}^{2}$. For large values of $x$ we observe an
underestimation of the data which is closely connected to the limit
$F_{2}^{n}/F_{2}^{p}\rightarrow 2/3$ as $x\rightarrow 1$. This is a well
known deficiency of the $SU(6)$--wave function we used at the bag level.
There are attempts to cure this problem \cite{SST91}, but one must
introduce at least two additional phenomenological
parameters at this point,
namely the diquark masses of the
intermediate spin singlet and spin triplet
states. Since we are not interested
in the limit $x\rightarrow 1$ anyhow,
we do not make modifications
in this direction. For small $x$ the data are
overestimated and we get, due to Eqs.\ (\ref{Norma}) and (\ref{Eq34}),
the value $1/3$ for the Gottfried sum
rule $\mbox{GSR}\equiv \int_{0}^{1}
\frac{dx}{x}(F_{2}^{p}(x,Q^{2})-F_{2}^{n}(x,Q^{2}))$. An excellent
description of the low--$x$ data and a reproduction of the experimental
GSR \cite{GSR1,GSR2} can be achieved
according to (\ref{Eq48}) by setting
$\phi^{u/U}\neq \phi^{u/D}$.
This requires, however, the introduction of
two parameters, namely $b^{u/U}$ and $b^{u/D}$, instead of just one
parameter $b$. Since we have not
found a {\em microscopic mechanism\/} yet which
could give rise to a difference
between $\phi^{u/U}$ and $\phi^{u/D}$, i.e. an $SU(2)$--breaking
at the level of the partons within the constituent quarks, we will not
study the GSR in this context, but
defer it to an extended version of the
convolution model which in addition to the three constituent quarks also
takes into account meson cloud effects.

\section{The polarized structure function $g_{1}$}

The investigations of Sect.\ 5 revealed that the convolution model
approach enables a satisfactory and consistent description of the
unpolarized structure functions, both as a function of $x$ and $Q^{2}$.
In the following we are going to apply the formalism
to the spindependent
structure function $g_{1}$ which is
given by the parton model representation
(\ref{Eq17}). Writing the polarized quark distributions as
$\Delta\phi_{v}={\cal P}_{v}\phi_{v}$ and
$\Delta\phi_{s}={\cal P}_{s}\phi_{v}$, with
\begin{equation} \label{Eq48a}
{\cal P}_{v,s}(z,Q^{2}) =
\frac{\phi_{v,s}^{\uparrow}(z,Q^{2})-
       \phi_{v,s}^{\downarrow}(z,Q^{2}) }
     {\phi_{v,s}^{\uparrow}(z,Q^{2})+
     \phi_{v,s}^{\downarrow}(z,Q^{2}) }\;,
\end{equation}
employing (\ref{Eq29}) and making use of $SU(6)$--symmetry at the
constituent quark level, we find the convolution model relation
\begin{eqnarray}
g_{1}^{p,n}(x,Q^{2}) & = & \int\limits_{x}^{1} \frac{dy}{y}
\biggl( g_{1}^{p,n}(y)\biggr)_{\scriptstyle\rm bag}
{\cal P}_{v} \!\!\left( \frac{x}{y},Q^{2} \right)\,
\phi_{v} \!\!\left( \frac{x}{y},Q^{2} \right) \nonumber \\
& & + \frac{11}{5}
\int\limits_{x}^{1} \frac{dy}{y}
\biggl( g_{1}^{p}(y)\biggr)_{\scriptstyle\rm bag}
{\cal P}_{s} \!\!\left( \frac{x}{y},Q^{2} \right)\,
\phi_{s} \!\!\left( \frac{x}{y},Q^{2} \right) \;. \label{Eq49}
\end{eqnarray}
This equation also implies the in our opinion reasonable assumption that
due to the comparatively small mass of the strange--quarks,
$m_{s}^{2}/Q^{2}\ll 1$, their polarization should not be considerably
suppressed compared to the polarization
of the u-- and d--quark flavours.
For reasons of simplicity we assume the same polarization for all three
flavours. Nonperturbative
effects enter the structure functions again via
the bag model, Eq.\ (\ref{Eq25}),
the functions $\phi$ are the same as before
and the spindependent information
is fully contained in the polarization
functions ${\cal P}_{v}$ and ${\cal P}_{s}$.

The particular form of (\ref{Eq49}) is perfectly appropriate to
perform a careful analysis of the
latest data. First of all we have a look
at ${\cal P}_{v}$. Imposing the fundamental Bjorken sum rule constraint
\cite{Bjo66}, modified by the leading order QCD--correction, we obtain
\begin{eqnarray}
\frac{1}{6}\left( \frac{g_{A}}{g_{V}} \right)
\left( 1-\frac{\alpha_{s}(Q^{2})}{\pi} \right) & = &
\int\limits_{0}^{1}dx\left( g_{1}^{p}(x,Q^{2})
                          - g_{1}^{n}(x,Q^{2}) \right)
\nonumber \\
& = &
\int\limits_{0}^{1} dy \,
\biggl( g_{1}^{p}(y)\biggr)_{\scriptstyle\rm bag}\:
\int\limits_{0}^{1} dz \,
{\cal P}_{v} ( z,Q^{2} )\,
\phi_{v}( z,Q^{2} ) \;, \label{Eq50}
\end{eqnarray}
so that
\begin{equation} \label{Eq51}
\int\limits_{0}^{1} dz \,
{\cal P}_{v} ( z,Q^{2} )\,
\phi_{v} ( z,Q^{2} ) =
\left( 1-\frac{\alpha_{s}(Q^{2})}{\pi} \right)\;,
\end{equation}
employing (\ref{Normb}). This relation
together with the normalization
(\ref{Eq34}) and the
condition $|{\cal P}_{v} (z,Q^{2})| \leq 1$ essentially
determines the shape of the
polarization function of the valence quarks.
Following Carlitz and Kaur \cite{CK77} we choose the parametrization
\begin{equation} \label{Eq52}
{\cal P}_{v} ( z,Q_{0}^{2} ) =
\left[ 1+ P_{v}\,\frac{(1-z)^{2}}{\sqrt{z}} \right]^{-1}
\end{equation}
at a fixed resolution scale $Q_{0}^{2} = 10\,\mbox{GeV}^{2}$.
In order to fulfill (\ref{Eq51}), $P_{v}$
has to be settled at $0.035$.

We now proceed to an investigation of the sea polarization.
As a starting
point it is rather instructive
to see which shape of $g_{1}^{p}$ would be
produced under the assumption of an entirely unpolarized sea, i.e.
${\cal P}_{s}(z,Q_{0}^{2})\equiv 0$.
The corresponding result is depicted
in Fig.\ 9. It is quite conspicuous that such a scenario leads to
a sizable overestimation of $g_{1}^{p}$ in the range
$0.01\leq x \leq 0.2$ where the sea plays an important role.
This observation suggests a negative sea polarization in this range
which we want to determine in the following.
In order to achieve this goal,
we try to reproduce the data as well as possible with the help of a
parametrization for ${\cal P}_{s}$
based on an analogy to Eq.\ (\ref{Eq52}).
For this purpose we find the choice
\begin{equation} \label{Eq53}
{\cal P}_{s} ( z,Q_{0}^{2} ) =
- \left[ 1+ P_{s}\,\frac{(1-z)^{2}}{z} \right]^{-1}
\end{equation}
to be fairly appropriate. Fig.\ 10 shows the sea contribution
to the polarized
structure function $g_{1}^{p}$ which is, combined with
the previously
discussed valence contribution, necessary to describe the
available data. This analysis can be directly transformed into the
polarization function of the
sea quarks, resulting in $P_{s} \approx 0.25$.
The reliability of this parametrization is, however,
restricted to the range
$0.01\leq x \leq 0.3$ since on the one hand
the influence of the sea quarks
is absolutely negligible for larger $x$ so that
in the range $x \geq 0.3$
the sea polarization
cannot be extracted from the data anyway. Moreover,
in the region $x\leq 0.01$ the SMC--data
feature a rapid increase and even lie above the valence
part of $g_{1}^{p}$, thus indicating that there may be a sign flip
in the polarization of the sea quarks which would cause a breakdown
of the parametrization (\ref{Eq53}).
Due to the huge error bars in the
very small--$x$ region such a conclusion is, however, premature
and one has to admit that there is at present no way to
exactly determine the behaviour of the polarization function for
$x\leq 0.01$.
This is just the crucial range
for the evaluation of the integral over
$g_{1}^{p}$ which sensitively depends on the extrapolation to
$x=0$. Therefore we believe that an analysis of the
spin structure of the
nucleon purely based on the first moment of $g_{1}^{p}$ is rather
questionable.

Taking our parametrization serious
even in the limit $x\rightarrow 0$, we
obtain the integral
$\int_{0}^{1}dx\,g_{1}^{p}( x,Q_{0}^{2} ) = 0.140$ and for
the total quark spin content we find
\begin{eqnarray}
\Sigma (Q_{0}^{2} ) & \equiv & \sum_{i=1}^{2n_{f}}
\int\limits_{0}^{1} dx\,\Delta q_{i} ( x,Q_{0}^{2} )
\nonumber \\
& = & \frac{18}{5} \int\limits_{0}^{1} dy\,
\biggl( g_{1}^{p}(y)\biggr)_{\scriptstyle\rm bag}\,
\left( \int\limits_{0}^{1} dz \,
{\cal P}_{v} ( z,Q_{0}^{2} )
\, \phi_{v} ( z,Q_{0}^{2} ) \right. \nonumber \\
& & \hspace{3.2cm} \left. + 5
\int\limits_{0}^{1} dz \, {\cal P}_{s} ( z,Q_{0}^{2} )
\, \phi_{s} ( z,Q_{0}^{2} ) \right) \nonumber \\
& = & 0.60 - 0.22 = 0.38
\label{Eq54}
\end{eqnarray}
which is considerably higher than the original results
\cite{BEK88,Lip91}. According to the spindependent evolution equations
\cite{AP77} this decomposition remains
unchanged for all values of $Q^{2}$.
This implies the following spin structure of the
proton. In the low energy
regime we have the dynamics of
three bag quarks whose relativistic motion
intrinsically also involves orbital angular momenta so that the quark
spin contribution reduces from 1 to $0.60$ at this level. In a DIS
experiment the virtual photon resolves the valence and sea quarks inside
the constituent quarks. While the valence
quarks are essentially polarized
parallel to the constituent quarks, the
spin alignment of the sea quarks,
at least for $x\geq 0.01$,
causes a partial shielding of the spin carried
by the valence quarks. The extrapolation to the very small--$x$ regime,
i.e. $x\leq 0.01$, is, however,
absolutely not clear yet so that we do not
want to rush to conclusions.
At any rate we believe that our results for
$\int_{0}^{1} dx\,g_{1}^{p}(x,Q_{0}^{2})$ and $\Sigma$ give lower
bounds for these quantities.
Furthermore we will not specify the flavour
decomposition of $\Sigma$ since this clearly depends on our
{\em assumptions\/} for the distribution of the strange--quarks as well
as their polarization.

Of course, the formalism of the convolution model can also be applied to
the polarized structure function of the neutron, $g_{1}^{n}$, as well so
that a prediction can be made. Proceeding on the assumption of
$SU(6)$--symmetry, $g_{1}^{n}$ is identical with the sea contribution to
$g_{1}^{p}$ and thus given by the sea part of (\ref{Eq49}).
The data of SLAC \cite{SLAC93} and SMC \cite{SMC93} having been taken,
however,
at an average $Q^{2}$ of $2\,\mbox{GeV}^{2}$ and  $4.6\,\mbox{GeV}^{2}$,
respectively, a $Q^{2}$--evolution
of the spindependent quark distributions
which we parametrized at $Q_{0}^{2} = 10\,\mbox{GeV}^{2}$ is required.
Like in the unpolarized case the evolution equations can be expressed in
terms of the functions characterizing the substructure of the
constituent quarks, with the only difference that the moments involving
the sum of spin up and down polarization,
Eqs.\ (\ref{Eq42})--(\ref{Eq44}), have to be replaced
now by the respective
ones involving the difference.
In order to perform the $Q^{2}$--evolution
explicitly, also the
gluon polarization function is needed in addition to
the previously studied polarization functions of valence and sea quarks.
For lack of any better knowledge we
take the polarized gluon distribution
function from a perturbative QCD calculation which
we are going to discuss
elsewhere \cite{KH94}. The resulting shape of the structure function
$g_{1}^{n}( x,\bar{Q^{2}} = 3.3\,\mbox{GeV}^{2} )$
at an average
$Q^{2}$ between the SLAC-- and SMC--data set is presented in
Fig.\ 11, clearly demonstrating that our convolution model
prediction for $g_{1}^{n}$ is in accordance
with experiment. From that one
can draw the conclusion that the sea polarization based on the analysis
of $g_{1}^{p}$ is fully consistent with the data
of the neutron structure
function.

\section{The polarized structure function $g_{2}$}

A formal analysis in the framework of the
operator product expansion (OPE) reveals that in contrast to $g_{1}$
the second spindependent structure function $g_{2}$
also receives twist--3
contributions \cite{Jaf90,JX91}. Accordingly we write
\begin{equation} \label{Eq55}
g_{2}( x,Q^{2} ) =
g_{2}^{\scriptstyle\rm T2}( x,Q^{2} ) +
g_{2}^{\scriptstyle\rm T3}( x,Q^{2} )\;.
\end{equation}
The twist--2 contribution of $g_{2}$ can be directly reconstructed from
$g_{1}$, yielding \cite{WW77}
\begin{equation} \label{Eq56}
g_{2}^{\scriptstyle\rm T2}( x,Q^{2} ) =
\int\limits_{x}^{1}\frac{dy}{y}\,
g_{1}( y,Q^{2} ) - g_{1}( x,Q^{2} )\;,
\end{equation}
whereas the twist--3 contribution is a consequence of the quark--gluon
correlations in the nucleon and with that also a reflection of
confinement,
thus representing
very complicated physics. These twist--3 effects cannot
be neglected a priori and should occur in a model which incorporates
any confinement mechanism.
In the bag model they are relatively large and
stem from the influence of the
bag surface which simulates confinement in
a phenomenological way \cite{JX91}.
The structure function $g_{2}$ involving
such complicated physics, the parton model interpretation of
$g_{T}\equiv g_{1}+g_{2}$, see (\ref{Eq18}), is of course
rather questionable so that, being rigorous, not even a basis of
parametrization is available yet.

In order to get a first
impression of $g_{2}$, however, we take a pragmatic
point of view in the following and maintain the parton model relation
(\ref{Eq18}) as an exemplary ansatz.
According to (\ref{Eq30}) the bag model
result (\ref{Eq26}) is then
interpreted in terms of transversely polarized
constituent quark distributions which are not identical to the
longitudinally polarized
quark distributions entering $g_{1}$, but also
receive modifications
due to quark--gluon correlations, i.e. the simulation
of confinement. Starting from
the constituent quark distributions, one again
can make the transition to the parton level via the convolution model
approach. In
order to give an estimate of $g_{2}$, let us now assume that
the twist--3 effects
are fully parametized by the bag model calculation.
In this case the substructure of the constituent quarks does not change
compared to the longitudinal polarization and the parton distribution
functions can be taken over from the
analysis of $g_{1}$, so that $g_{T}$
is just determined by (\ref{Eq49}) after replacing
$(g_{1}(y))_{\scriptstyle\rm bag}$ by
$(g_{T}(y))_{\scriptstyle\rm bag}$. From
that we finally extract $g_{2}$.
The result is presented in Fig.\ 12 which separately displays
the twist--2 and twist--3 contributions.
The twist--3 part shows a very
similar characteristic in $x$
like the absolute value of the twist--2 part
and both are definitely of the
same order of magnitude. For the integral
over $g_{2}$ we find the value $0.002$ which is very close to the
Burkhardt--Cottingham sum rule
$\int_{0}^{1}dx\,g_{2}^{p}(x) = 0$ \cite{BC70}. There is
no evidence so far that this sum rule may be violated.

Concluding the considerations of
this section, we would like to emphasize
once more that our
previously given result of $g_{2}$ is not supposed to
be a high precision prediction,
but thought as a reasonable estimate based
on the convolution model. For lack of any feasible scheme for the full
$Q^{2}$--evolution of twist--3 structure functions
there is also no way to
predict $g_{2}$ at various $Q^{2}$, provided it is known at a scale
$Q_{0}^{2}$.

\section{Conclusions and outlook}

In this work we presented a
convolution model approach for the investigation
of the nucleon structure functions.
Such a model is based on the concept
of effective degrees of freedom
describing the low energy properties of the
nucleon. In DIS, however, their
fermionic substructure, characterized by
the parton distribution
functions $\phi_{v}$ and $\phi_{s}$, is directly
probed. These functions involve
three parameters which were determined at a
reference scale $Q_{0}^{2} = 10\,\mbox{GeV}^{2}$
by experimental input.
The $Q^{2}$--evolution equations
could be formulated at the level of the
parton distributions $\phi$, thus unravelling the substructure of the
constituent quarks in dependence on the resolution scale. Comparing our
results for $F_{2}^{p}(x,Q^{2})$ at various $Q^{2}$ with the data,
we found
the picture of the convolution
model to be fully consistent with experimental
findings. In the case of the
difference $F_{2}^{p}-F_{2}^{n}$ we observed,
however, obvious discrepancies
in the large-- as well as in the small--$x$
region which can be attributed to the use of $SU(6)$--symmetry at the
constituent quark
level and isospin symmetry of the sea quarks, respectively.

Subsequent to the investigation of the unpolarized structure functions
we applied the formalism of the convolution model to the spindependent
structure functions $g_{1}^{p,n}$. In addition to the nonperturbative
information of the bag model calculation and the distribution functions
$\phi$ the expressions for
$g_{1}^{p,n}$ also involve polarization functions of the partons. An
analysis of the $x$--dependence of the proton data taken so far implied
that the valence quarks are predominantly polarized parallel to the
constituent quarks while the sea quarks tend to screen the spin
contribution
of the valence quarks.
Having extrapolated our parametrization of the sea
polarization function, which can be pinned down rather well for
$0.01\leq x \leq 0.2$, to $x=0$, we obtained the quark spin
decomposition
$\Sigma=\Sigma_{v}+\Sigma_{s}=0.60-0.22=0.38$. We would like to
emphasize, however,
that the small $x$--behaviour of the sea polarization,
particularly the one for $x\leq 0.01$,
and accordingly also the integrated
spin content of the sea quarks remain
essentially undetermined yet. For this
reason one definitely cannot exclude the possibility of a totally
unpolarized sea, i.e. $\Sigma_{s}=0$, at the moment. With the
polarization functions in hands we could finally make a prediction for
$g_{1}^{n}$. Comparison with experiment indicated that the sea
contribution to the proton
structure function is absolutely consistent with
the full neutron structure function. From that
one can also conclude that the
proton and neutron data together are perfectly compatible with the
fundamental Bjorken sum rule which is fulfilled in our model per
construction.

In the last section of our work we presented an estimate for the
transversely polarized
structure function $g_{2}$. In doing this we took
the naive parton model
interpretation as a reasonable guideline and assumed
that the twist--3 effects are fully parametrized by the confinement
mechanism of the bag model.
The resulting estimate for $g_{2}^{p}$ in the
convolution model suggests
that twist--3 contributions may be relatively
important even
for values of $Q^{2}$ of the order of $10\,\mbox{GeV}^{2}$.

As became clear in the course of our discussions there are some points
which call for further investigations. The first problem is associated
with the small--$x$ region of $F_{2}^{p}-F_{2}^{n}$ and consequently
connected with the deviation of the GSR from $1/3$. Here we aim at a
microscopic mechanism which explains the preference of d--quarks over
u--quarks in the proton sea. This requires an extension of the
convolution
model which in addition to the three constituent quarks also includes a
meson cloud. Another challenge consists in achieving a microscopic
understanding of the sea polarization as given by our analysis. The spin
problem would then finally loose its magic.
Especially this issue is intensively worked on \cite{KH94}.

\newpage

\noindent
{\large\bf Figure captions}

\vspace{0.3cm}

\noindent
Fig.\ 1.  Deep inelastic lepton--nucleon scattering in the
          one photon exchange approximation.

\vspace{0.3cm}

\noindent
Fig.\ 2.  DIS in the convolution model for the nucleon substructure.

\vspace{0.3cm}

\noindent
Fig.\ 3.  The unpolarized structure function
          $F_{2}^{p}( x,Q_{0}^{2}=10\,\mbox{GeV}^{2} )$
          in the
          convolution model. The data are taken from \cite{BCDMS-F2}.

\vspace{0.3cm}

\noindent
Fig.\ 4.  $Q^{2}$--evolution of $\phi_{v}( z,Q^{2} )$.

\vspace{0.3cm}

\noindent
Fig.\ 5.  $Q^{2}$--evolution of $\phi_{s}( z,Q^{2} )$.

\vspace{0.3cm}

\noindent
Fig.\ 6.  $F_{2}^{p}$ evolved to $Q^{2}=5\,\mbox{GeV}^{2}$.
          The data are
          taken from \cite{NMC-F2}.

\vspace{0.3cm}

\noindent
Fig.\ 7.  $F_{2}^{p}$ evolved to $Q^{2}=50\,\mbox{GeV}^{2}$.
          The data are
          taken from \cite{BCDMS-F2}.

\vspace{0.3cm}

\noindent
Fig.\ 8.  $[ F_{2}^{p}-F_{2}^{n} ]
          ( x,Q^{2}=40\,\mbox{GeV}^{2} )$
          under the assumption
          of an isospin symmetric sea.
          The data are taken from \cite{GSR1}.

\vspace{0.3cm}

\noindent
Fig.\ 9.  $g_{1}^{p}( x,Q_{0}^{2}=10\,\mbox{GeV}^{2} )$
          in the convolution model under the
          assumption of an identically vanishing
          sea polarization. The data
          are taken from \cite{EMC89,SMC94,SLACg1p}.

\vspace{0.3cm}

\noindent
Fig.\ 10. Valence and sea contribution to the structure function
          $g_{1}^{p}( x,Q_{0}^{2}=10\,\mbox{GeV}^{2} )$.
          The data are taken from \cite{EMC89,SMC94,SLACg1p}.

\vspace{0.3cm}

\noindent
Fig.\ 11. The neutron structure function
          $g_{1}^{n}( x,\bar{Q^{2}} )$ at an average
          $\bar{Q^{2}}=3.3\,\mbox{GeV}^{2}$ in comparison with the
          data \cite{SLAC93,SMC93}.

\vspace{0.3cm}

\noindent
Fig.\ 12. An estimate for
          $g_{2}^{p}( x,Q_{0}^{2}=10\,\mbox{GeV}^{2} )$
          in the convolution model.


\begin{thebibliography}{99}

\bibitem{Cho74/1}A. Chodos et al.: Phys. Rev. D 9 (1974) 3470
%
\bibitem{Jaf75}  R.L. Jaffe: Phys. Rev. D 11 (1975) 1953
%
\bibitem{Hug77}  R.J. Hughes: Phys. Rev. D 16 (1977) 622
%
\bibitem{Bar79}  J.A. Bartelski: Phys. Rev. D 20 (1979) 1229
%
\bibitem{Alt74}  G. Altarelli et al.: Nucl. Phys. B 69 (1974) 531
%
\bibitem{Hwa80}  R.C. Hwa: Phys. Rev. D 22 (1980) 759
%
\bibitem{EMC89}  J. Ashman et al., EM Coll.: Nucl. Phys. B 328 (1989) 1
%
\bibitem{SMC94}  D. Adams et al., SM Coll.: Phys. Lett. B 329 (1994) 399
%
\bibitem{SLAC93} P.L. Anthony et al., E--142 Coll.: Phys. Rev. Lett. 71
                 (1993) 959
%
\bibitem{SMC93}  B. Adeva et al., SM Coll.: Phys. Lett. B 302 (1993) 533
%
\bibitem{Prop1}  J. Beaufays et al., SM Coll.:
                 CERN/SPSC 88--47, SPSC/P242
                 (1988)
%
\bibitem{Prop2}  R. Arnold: SLAC Proposal E--142 (1989)
%
\bibitem{Jaf90}  R.L. Jaffe: Comm. Nucl. Part. Phys. 19 (1990) 239
%
\bibitem{Clo79}  F.E. Close: An introduction to quarks and partons.
                 New York: Academic Press 1979
%
\bibitem{CL88}   T. Cheng, L. Li: Gauge theory of elementary
                 particle physics. Oxford: Clarendon Press 1988
%
\bibitem{Fey72}  R.P. Feynman: Photon hadron interactions. New York:
                 Benjamin 1972
%
\bibitem{AP77}   G. Altarelli, G. Parisi: Nucl. Phys. B 126 (1977) 298
%
\bibitem{JX91}   R.L. Jaffe, X. Ji: Phys. Rev. D 43 (1991) 724
%
\bibitem{Cho74/2}A. Chodos et al.: Phys. Rev. D 10 (1974) 2599
%
\bibitem{BM87}   C.J. Benesh, G.A. Miller: Phys. Rev. D 36 (1987) 1344
%
\bibitem{SST91}  A.W. Schreiber, A.I. Signal, A.W. Thomas:
                 Phys. Rev. D 44 (1991) 2653
%
\bibitem{PY57}   R.E. Peierls, J. Yoccoz: Proc. Phys. Soc. A 70 (1957) 381
%
\bibitem{Jaf85}  R.L. Jaffe in: Relativistic dynamics and
                 quark--nuclear physics,
                 M.B. Johnson, A. Picklesimer (Eds.):
                 Proc. of the 1985
                 Los Alamos Workshop. New York: Wiley 1985
%
\bibitem{Dzi89}  Z. Dziembowski et al.: Phys. Rev. D 39 (1989) 3257
%
\bibitem{Rob90}  R.G. Roberts: The structure of the proton. Cambridge:
                 Cambridge University Press 1990
%
\bibitem{BCDMS-F2}A.C. Benvenuti et al., BCDMS Coll.: Phys. Lett. B 223
                 (1989) 485
%
\bibitem{Bur80}  A. Buras: Rev. Mod. Phys. 52 (1980) 199
%
\bibitem{NMC-F2} P. Amaudruz et al., NM Coll.:
                 Phys. Lett. B 295 (1992) 159
%
\bibitem{GSR1}   A.C. Benvenuti et al., BCDMS Coll.: Phys. Lett. B 237
                 (1990) 599
%
\bibitem{GSR2}   P. Amaudruz et al., NM Coll.: Phys. Rev. Lett. 66
                 (1991) 2712; CERN--PPE/93--117 (1993)
%
\bibitem{Bjo66}  J.D. Bjorken: Phys. Rev. 148 (1966) 1467
%
\bibitem{CK77}   R. Carlitz, J. Kaur: Phys. Rev. Lett. 38 (1977) 673
%
\bibitem{SLACg1p}M.J. Alguard et al., SLAC--Yale Coll.:
                 Phys. Rev. Lett. 37
                 (1976) 1261;
                 G. Baum et al., SLAC--Yale Coll.: Phys. Rev. Lett. 45
                 (1980) 2000; 51 (1983) 1135
%
\bibitem{BEK88}  S.J. Brodsky, J. Ellis, M. Karliner:
                 Phys. Lett. B 206 (1988) 309
%
\bibitem{Lip91}  H.J. Lipkin: Phys. Lett. B 256 (1991) 284
%
\bibitem{KH94}   J. Keppler, H.M. Hofmann: work in progress
%
\bibitem{WW77}   S. Wandzura, F. Wilczek: Phys. Lett. B 72 (1977) 195
%
\bibitem{BC70}   H. Burkhardt, W.N. Cottingham: Ann. Phys. 56 (1970) 453

\end{thebibliography}
\end{document}